# Expanding Accessibility in Immersive Virtual Spaces: A Comprehensive Approach for All Disabilities

Cecilia Aragon

University of Washington, Seattle, WA, USA, aragon@uw.edu

Melissa Vosen Callens

North Dakota State University, Fargo, ND, USA, melissa.vosen@ndsu.edu

Stacy M. Branham

University of California, Irvine, Irvine, California, USA, sbranham@uci.edu

Cali Anicha

North Dakota State University, Fargo, ND, USA, calianicha@gmail.com

Brianna Blaser

University of Washington, Seattle, WA, USA, blaser@uw.edu

Canan Bilen-Green

North Dakota State University, Fargo, ND, USA, canan.bilen.green@ndsu.edu

In the early stages of the COVID-19 pandemic, many events and conferences hastily converted to a virtual format, and many commercial ventures promptly developed tools promising seamless transitions to virtual spaces. In particular, efforts to expand and monetize augmented and virtual reality environments increased. While these spaces increased accessibility for some, others were left behind. In 2024, many events returned to on-site venues, yet virtual spaces remain central in academic and research communities, particularly for disabled scholars. As such, in this paper, we advocate for continued virtual access and improved virtual spaces; we also identify some potentially overlooked harms in immersive and embodied virtual spaces.

## 1 INTRODUCTION

At the beginning of the COVID-19 pandemic, conferences and other events developed significant interest in transitioning to virtual spaces, including augmented and virtual reality environments. As just a few examples, CHI 2021 adopted Delegate Connect, IMX 2021 adopted OhYay, and CSCW 2022 adopted Gathertown. During this rush of innovation, there was some attention paid to the way that many on-site experiences are exclusionary to blind/low-vision people, d/Deaf/hard-of-hearing people, and people with invisible disabilities, physical disabilities, or mental health conditions. For those in the disability community, there was a brief moment of hope that the pandemic was facilitating a long overdue reckoning with conference inaccessibility. Unfortunately, this was not the case. In the vendor stampede to monetize the changes needed to gather safely during a pandemic, many tools were branded "accessible" without careful attention to limitations. For example, CSCW 2023's platform, Midspace, claimed the ability to provide auto-captions in 2022 [1], yet was unable to do so by the time the conference was held in November 2023. Other vendors ignored accessibility outright because the demand for their product was high. In reality, a significant portion of the population, including individuals with disabilities, were being denied access to the current generation of augmented and virtual reality technology [2]. The authors of this paper are a group of academics who have been working in the field of disability education for many years, and some of us also have disabilities. As such, we wish to amplify overlooked harms in immersive and embodied virtual spaces and offer suggestions to mitigate these harms. Examples of groups who were excluded from full participation in virtual spaces follow, such as people who rely on voice recognition software to access a virtual environment, followed by recommendations for product teams and event organizers.

## 2 COMPREHENSIVE ACCESSIBILITY

We advocate for a holistic approach to the development of virtual spaces, one that considers all types of disabilities. We term this "comprehensive accessibility." For example, a virtual space should consider the importance of luminance contrast for low-vision users [3], as well as the use of specific palettes that are suitable for individuals with color vision deficiency. It should also consider that some users cannot see the screen and therefore need complete access through the keyboard and screen reader output; though, there should not be an assumption that all users can click a mouse or type on a keyboard. Furthermore, people who struggle with conditions related to vertigo may not do well in immersive environments. People who are autistic or who have attention-deficit/hyperactivity disorder (ADHD) may find the social cues in virtual environments distracting and overwhelming [4]. Finally, there may be accessibility concerns related to social anxiety and how to appropriately interact with the technology and others [5]. These are only six examples that people with disabilities may experience that often are not considered by vendors or organizations. There should be easy options for people with all disabilities to access immersive spaces, not merely an "accessible" label on tools that require dozens of hours of customization on the part of disabled users. Additionally, user-centered documentation and tutorials should be readily available to orient all users to the environment [6].

## 3   USER-CENTERED DESIGN

It is vital to include individuals with multiple types of disabilities in all stages of the design process when creating accessible virtual spaces. The product team—engineers, UX designers, content designers, testers—should be diverse and include disabled individuals; in other words, accessibility is everyone's responsibility [7]. User research participants should include users from a variety of backgrounds, such as those who can't click a mouse, have vertigo, or have anxiety issues, reflecting users with different types of disabilities and different lived experiences. Because technology companies often don't have sufficient accessibility staff, they may make false claims about accessibility compliance. It is not uncommon for voluntary product accessibility templates to contain false or misleading information [8]. Other companies rely upon accessibility overlays that are focused exclusively on one or two types of disability and that don't actually lead to fully accessible technology.

## 4   TECHNOLOGY AND INNOVATION

The excitement of augmented and virtual reality, and the increasing power of computational engines, run the risk of a technology-led practice rather than a user-centric one. The "move fast and break things" ethos has been shown to produce significant amounts of harm, particularly to marginalized or vulnerable groups. In our own personal experience, authors of this paper have seen event organizers spend a great deal of money on so-called technological advancements, often without proper consideration of the diverse populations the technology intends to serve. Product design and development teams need to be wary of any unintended consequences of their design, especially how the product may inadvertently limit participation in some populations. We additionally recommend event organizers follow the guidance in ACCESS SIGCHI's Accessible Virtual Conference Guide [9], testing vendor technologies rather than trusting dubious claims of accessibility.

## 5   LEGAL AND ETHICAL CONSIDERATIONS AND RECOMMENDATIONS

Both legally and ethically, it's the responsibility of designers and developers to create inclusive virtual environments. Additionally, they are obligated to provide accurate descriptions of their products' capabilities and refrain from misleading users. Moreover, as we have repeatedly stressed, it is crucial to adhere to user-centered design principles when developing augmented reality and virtual reality environments. Conference and meeting organizers should put in place processes to ensure their event is accessible; this includes everything from the technology employed to having an accessibility chair. Organizers should ask about accessibility in event evaluations and include accessibility and disability in any DEIA activities/committees/etc.

## 6   CONCLUSION

Augmented and virtual reality environments are often lauded as offering higher degrees of presence as opposed to other mediated environments. Still, we must ask: Immersive for whom? Without the involvement of disabled users in the design process, these environments can be even more restrictive and less accessible than older media technologies. Ultimately, an inaccessible product leads to poor user experiences, and, when they do (particularly in the case of "new" technologies), large groups and organizations often abandon their use for what they know—on-site events with minimal technology, which are problematic in their own right. Virtual spaces remain central in academic and research communities, particularly for disabled scholars, making them critical. It is in everyone's best interest to invest time, money, and energy into making these spaces accessible. When we can all participate fully, the overall user experience improves for everyone.